\documentclass[aps,twocolumn,english,superscriptaddress,citeautoscript,preprintnumbers,amsmath,amssymb,floatfix,footinbib]{revtex4-2}
\usepackage[breaklinks=true,colorlinks,citecolor=blue,linkcolor=blue,urlcolor=blue]{hyperref}
\usepackage{amsmath}
\usepackage{graphicx}
\usepackage{amssymb}
\usepackage{dcolumn}
\usepackage{float}
\usepackage[utf8]{inputenc}
\DeclareUnicodeCharacter{2212}{-}
\usepackage{color}
\usepackage[super]{nth}
\begin{document}
	
\title{Multiple magnetic transitions, metamagnetism and large magnetoresistance in GdAuGe single crystals}
\author{D. Ram}
\affiliation{Department of Physics, Indian Institute of Technology, Kanpur 208016, India}
\author{J. Singh}
\affiliation{Department of Physics, Indian Institute of Technology Hyderabad, Kandi, Medak 502 285, Telangana, India}
\author{ M. K. Hooda}
\affiliation{Department of Physics, Indian Institute of Technology, Kanpur 208016, India}
\author{K. Singh}
\affiliation{\mbox{Institute of Low Temperature and Structure Research, Polish Academy of Sciences, ul. Okolna 2, 50-422 Wroclaw, Poland}}
\author{V. Kanchana}
\email{kanchana@iith.ac.in}
\affiliation{Department of Physics, Indian Institute of Technology Hyderabad, Kandi, Medak 502 285, Telangana, India}
\author{D. Kaczorowski}
\email{d.kaczorowski@intibs.pl}
\affiliation{\mbox{Institute of Low Temperature and Structure Research, Polish Academy of Sciences, ul. Okolna 2, 50-422 Wroclaw, Poland}}
\author{Z. Hossain}
\email{zakir@iitk.ac.in}
\affiliation{Department of Physics, Indian Institute of Technology, Kanpur 208016, India}

\begin{abstract}
We report the physical properties of GdAuGe single crystals, which were grown using Bi flux. The powder x-ray diffraction data shows that the compound crystallizes in hexagonal NdPtSb-type structure (space group \textit{P6}$_3$\textit{mc}). Magnetization measurements performed for field configuration \mbox{\textit{H} $\parallel$ \textit{c}} and \textit{H} $\perp$ \textit{c} show that GdAuGe orders antiferromagnetically at the N\'{e}el temperature, \textit{T}$_N$ = 17.2 K. Around this temperature, heat capacity and electrical resistivity data exhibit prominent anomaly due to the antiferromagnetic (AFM) transition. In addition to an AFM phase transition, the magnetization data for \textit{H} $\parallel$ \textit{c} display the signature of field-induced metamagnetic (MM) transitions below \textit{T}$_N$. The critical field range for these transitions vary from 0.2 to 6.2 T. The critical fields for the MM transitions decrease with increasing temperature and approach to zero value for temperature approaching \textit{T}$_N$. For instance, in high field MM transition, critical field changes from 6.2 T at 1.7 K to 1.8 T at 16 K. Interestingly, the magnetoresistance (MR) data (for \textit{H} $\parallel$ \textit{c}) record a sharp increase in values at the critical fields that coincide with those seen in magnetization data, tracking the presence of MM transitions. MR is positive and large ($\approx$ 169\% at 9 T and 2 K) at low temperatures. Above \textit{T}$_N$, MR becomes small and switches to negative values. Hall resistivity data reveal the predominance of hole charge carriers in the system. In addition, we observe an emergence of step-like feature in the Hall resistivity data within the field range of second MM, and a significantly large anomalous Hall conductivity of $\sim$ 1270 $\Omega$$^{-1}$ cm$^{-1}$ at 2 K. The \textit{H}$-$\textit{T} phase diagram constructed from our detailed magnetization and magnetotransport measurements reveals multiple intricate magnetic phase transitions. The electronic and magnetic structure of GdAuGe are also thoroughly investigated using first-principles methods. The electronic band structure calculations reveal that GdAuGe is a Dirac nodal-line semimetal.
	
\end{abstract}

\maketitle
\section{Introduction}
Ternary rare-earth intermetallic compounds continue to receive the attention of scientific community because of their complex relationships between composition and structure, interesting magnetic, thermodynamic and transport properties \cite{RTX_1999,REAuTt,RAuGe,NdAuGe,RAgGe_2004,NdAuGe_2008}. These compounds possess a wide range of magnetic characteristics, ranging from simple diamagnetic behavior to very complex magnetic phases, depending upon the degree of hybridization between 4\textit{f} and conduction electrons \cite{RAgGe_2004}. The majority of these compounds either display antiferromagnetic (AFM) or ferromagnetic (FM) ordering due to the long-range nature of dominant Ruderman-Kittel-Kasuya-Yosida interactions present in them \cite{RAgGe_2004,NdAuGe_2008,RPtIn,RareM,EuAuAs}. The application of an external magnetic field to the AFM ground states of some rare-earth compounds is observed to disrupt their low magnetization state, leading to metamagnetic (MM) transitions \cite{RAgGe_2004,Metamagnetism,RAgSi}. These transitions occur in both strongly and weakly anisotropic magnetic structures and are very sensitive to the crystalline electric field (CEF) effects, which constrain the magnetic moments along a specific axis \cite{RAgGe_2004,RPtIn,Metamagnetism,RAgSi,RGaGe}. For example, GdAgSi exhibits one MM transition at 4 K with critical field of $\sim$ 0.29 T \cite{RAgSi}.

Furthermore, these intermetallic compounds become more interesting, when interplay between magnetism and novel electronic states generates new exotic quantum states and intriguing physical properties such as quantum critical behavior \cite{Quantum_Nature,Quantum_RMP}, and unconventional superconductivity \cite{Eu(FeIr)2As2,RMP_Superconducting}. Among them, Eu- and Gd-based intermetallic compounds attract more attention due to their oxidation state Gd$^{3+}$ and Eu$^{2+}$ (most stable oxidation state) having the electron configuration 4\textit{f}$^7$ with a half-filled \textit{f} shell, resulting in a quenched orbital momentum and very weak spin-orbit coupling (SOC). The \textit{f}-electron systems having magnetic frustration can also give rise to a skyrmion phase or noncollinear spin texture with nonzero scalar spin chirality [\mbox{$\chi_s$ = \textit{S}$_i$·(\textit{S}$_j$$\times$\textit{S}$_k$) $\ne$ 0}, where \textit{S}$_i$, \textit{S}$_j$, and \textit{S}$_k$ are the three nearest spins]. These can act as a fictitious magnetic field on the conduction electrons, giving rise to the topological Hall effect (THE) \cite{THE2012,MnPdGa,Y2Mo2O7}. For example, Gd$_2$PdSi$_3$ is a centrosymmetric triangular lattice with AFM ordering, it exhibits an intrinsic THE arising from a skyrmion phase under magnetic field \cite{Gd2PdSi3}. It is further interesting to note that GdAgGe single crystals previously investigated by us do not show any MM character up to field of 7 T, but it is a topological nodal line containing the drumhead surface states in \textit{k}$_z$ = 0 plane, protected by the inversion symmetry \cite{GdAgGe2023}. Thus, it becomes important to examine whether it is possible to induce MM transitions and novel electronic state in rare-earth based germanide systems by substituting transition metals.

In this context, equiatomic rare-earth gold-germanide (RAuGe) series can be interesting. The compounds of this series crystallize in the non-centrosymmetric hexagonal crystal structure with space group \textit{P6}$_3$\textit{mc}, where two-dimensional infinite chains of [AuGe] polyanions are separated by rare earth element ions \cite{RAuGe}. The detail studies on the magnetic and physical properties of polycrystalline RAuGe compounds have been already reported in the literature \cite{RAuGe,NdAuGe,HoAuGe,Er_HoAuGe,RAuGe1999,GdAuGe1996}. It is observed that HoAuGe and NdAuGe in RAuGe series display the MM transitions at 2 K with the critical fields of 0.4 and 3 T, respectively \cite{NdAuGe,HoAuGe}. Here, we focus on another member of RAuGe series i.e. GdAuGe. Polycrystalline GdAuGe was reported to order antiferromagnetically at 16.9 K \cite{GdAuGe1996,GdAgGe1998}. 

In this report, we study the anisotropic magnetic and electronic transport properties of GdAuGe single crystals with high magnetic fields up to 9 T, as well as detailed electronic structure using first-principles calculations. Our study on GdAuGe single crystals shows the AFM ground state at 17.2 K for fields perpendicular and parallel to the crystallographic \textit{c} axis. The field applied parallel to the \textit{c} axis of crystal induces MM transitions below the AFM ordering temperature, which correlate well with the magnetotransport properties of the compound. Furthermore, we present a first-principles calculations of the magnetic ground state and topological character of the compound.

\section{Experimental details}

\begin{figure}
	\includegraphics[width=8.7cm, keepaspectratio]{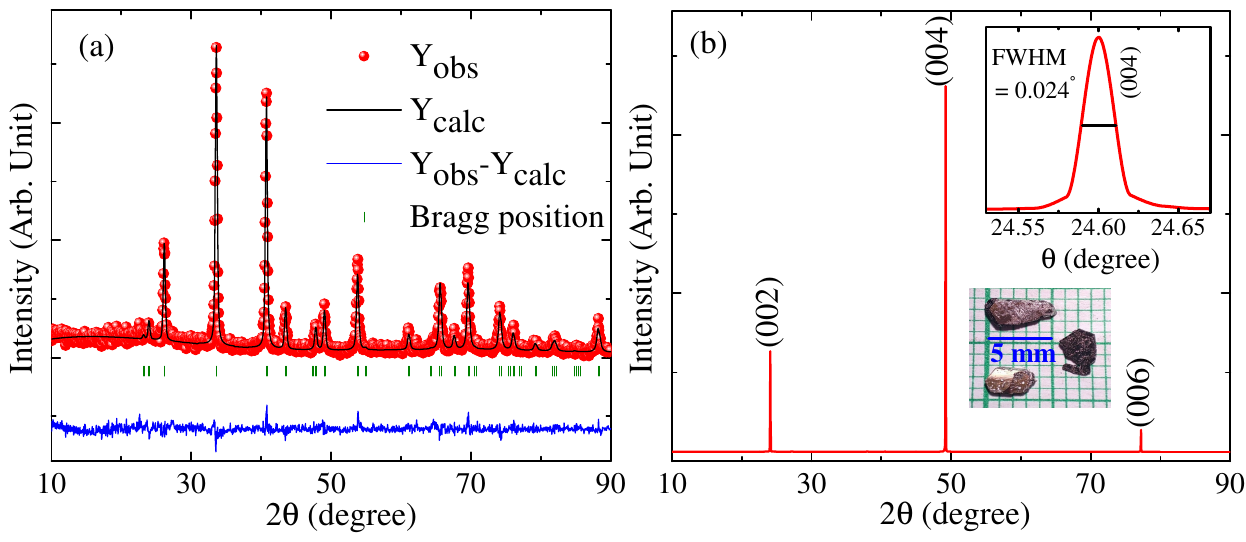}
	\caption{\label{XRD}(a) Rietveld refined powder XRD patterns of crushed single crystals of GdAuGe at room temperature. The blue line represents the difference between the observed intensity (red solid circles) and the calculated intensity (solid black line). The olive vertical lines represent the position of Bragg peaks. (b) The single-crystal XRD pattern of a GdAuGe, showing only (00$l$) reflections. Upper inset shows the rocking curve of peak (004). Lower inset shows a optical image of crystals.}
\end{figure}

The single crystals of GdAuGe were synthesized using Bi as an external flux. Starting elements Gd (99.9\%, Alfa Aesar), Au (99.99\%, Alfa Aesar), Ge (99.999\%, Alfa Aesar) and Bi (99.999\%, Alfa Aesar) were taken in a molar ratio of 1:1:1:10. The constituent elements were put into an alumina crucible, which was then transferred to a silica quartz tube. The tube was sealed under partial pressure of argon gas. In the next step, the sealed assembly was heated to temperature of 1050 $^\circ$C, where it was held for 24 h in order to obtain homogeneous solution. Subsequently, the slow cooling to 680 $^\circ$C at the rate of 2.5 $^\circ$C/h produced very shiny plate-like single crystals with a typical size of 5 $\times$ 3 $\times$ 0.4 mm$^3$ (as shown in bottom inset of Fig. \ref{XRD}(b)), which were separated from the Bi flux by centrifuging. 

The phase purity and orientation of as grown crystals were analyzed by x-ray diffraction (XRD) using a PANalytical X’Pert PRO diffractometer with Cu K$_{\alpha1}$ radiation. The XRD pattern of powdered crystals and a representative single crystal recorded at room temperature is shown in Figs. \ref{XRD}(a) and \ref{XRD}(b), respectively. It confirms the single phase growth of the compound crystallizing in the hexagonal crystal structure with space group \textit{P6}$_3$\textit{mc} (No. 186). The lattice parameters ($a$ = $b$ = 4.4281 \AA,  and $c$ = 7.4262  \AA) obtained from Rietveld refinement are in good agreement with previously reported data in the literature \cite{GdAuGe1996,GdAgGe1998,RAuGe2023}. The presence of (00$l$) peaks in the single crystal diffraction pattern shows that the crystallographic $c$ axis of crystal is perpendicular to its flat plane. The upper inset of Fig. \ref{XRD}(b) presents the rocking curve of (004) peak with a full width at half maximum (FWHM) $\Delta$$\theta$ = 0.024$^{\circ}$, indicating a high quality of the single crystal used. The desired chemical composition of crystals was further confirmed by energy-dispersive x-ray spectroscopy using a JEOL JSM-6010LA scanning electron microscope. Electrical resistivity and magnetoresistance measurements were performed using a Quantum Design physical property measurement system (PPMS) by the standard four-probe method. Heat capacity measurements were performed by the conventional relaxation method in the same PPMS platform. The magnetic susceptibility and magnetization were measured down to 1.7 K using a Quantum Design magnetic property measurement system.

Based on the density functional theory (DFT) \cite{DFT1, DFT2}, the first-principles calculations were carried out using the projector augmented wave \cite{PAW} approach as implemented in the Vienna ab initio simulation package \cite{VASP1, VASP2}. The generalised gradient approximation (GGA) with the Perdew-Burke-Ernzerhof \cite{PBE} parametrization was utilised to account for exchange-correlation effects. A Hubbard U parameter (GGA+U) of 6 eV was used to address the correlation effects of Gd-$f$ states \cite{U1, U2}. The calculations were done with a plane wave energy cutoff of 600 eV, and the energy convergence criterion was chosen to be 10$^{-8}$ eV. The geometry optimization was performed with 2 $\times$ 2 $\times$ 2 supercell using a 16 $\times$ 16 $\times$ 8 \textit{k}-mesh as per the Monkhorst-Pack method \cite{Monkhorst}.

\begin{figure}
	\includegraphics[width=8.70cm, keepaspectratio]{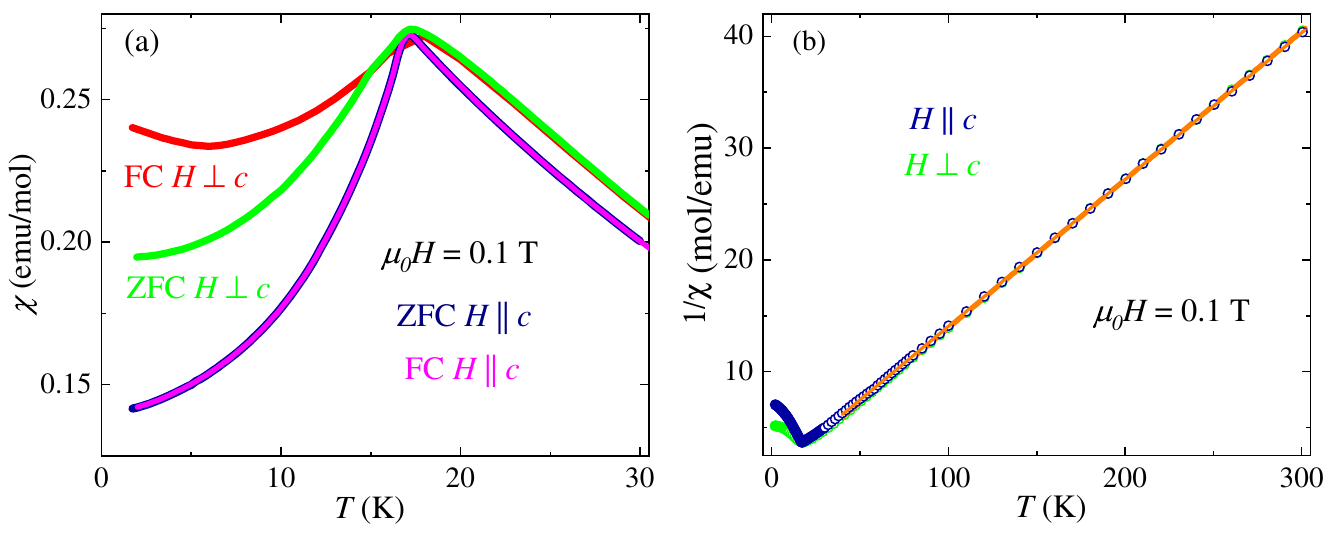}
	\caption{\label{MT}(a) Temperature-dependent magnetic susceptibility of GdAuGe measured under an applied magnetic field of $\mu_0$\textit{H} = 0.1 T for \textit{H} $\parallel$ \textit{c} and \textit{H} $\perp$ \textit{c} in ZFC and FC modes. (b) The inverse magnetic susceptibility as a function of temperature for \textit{H} $\parallel$ \textit{c} and \textit{H} $\perp$ \textit{c}. The solid orange lines show the Curie-Weiss fit above 50 K.}
\end{figure}

\section{Results and discussion}

\subsection{Magnetic properties}

Figure \ref{MT}(a) presents the temperature (\textit{T}) dependence of magnetic susceptibility ($\chi$) measured under zero-field cooling (ZFC) and field cooling (FC) conditions at the constant magnetic field of 0.1 T applied perpendicular and parallel to the crystallographic \textit{c} axis. A maximum of $\chi$(\textit{T}) is visible at \textit{T}$_{N}$ = 17.2 K for both field configurations, which is indicative of an AFM ordering in the compound and marks the boundary between AFM and paramagnetic (PM) phase. This value is very close to the previously reported \textit{T}$_{N}$ for GdAuGe \cite{GdAuGe1996,GdAgGe1998,RAuGe2023}. It is to be noted that $\chi$(\textit{T}) shows bifurcation in ZFC$-$FC measurements below 15 K for field configuration \textit{H} $\perp$ \textit{c}. It points out spin reorientations in the compound below 15 K, which also corroborates with second anomaly in heat capacity data. The magnetic anisotropy of the system in the AFM region is low as evident from Fig. \ref{MT}(b). It becomes insignificant in the PM region (above \textit{T}$_{N}$). Above 50 K, the data plotted as inverse magnetic susceptibility ($\chi^{-1}$) vs. \textit{T} in Fig. \ref{MT}(b) fit to the Curie-Weiss formula $\chi(T) = C/(T-\Theta)$, where $C$ and $\Theta$ are the Curie constant and Curie-Weiss temperature, respectively. The least-square fitting yields the value of $\Theta$ $\approx$ -4.7 and -6.9 K for \textit{H} $\perp$ \textit{c} and \textit{H} $\parallel$ \textit{c}, respectively. The estimated effective magnetic moment of $\mu_{eff}$ = 7.76 (for \textit{H} $\perp$ \textit{c}) and 7.80 $\mu_B$/Gd (for \textit{H} $\parallel$ \textit{c}) is in close agreement with the theoretical value expected for Gd$^{3+}$ ion.

\begin{figure}
	\includegraphics[width=7.0cm, keepaspectratio]{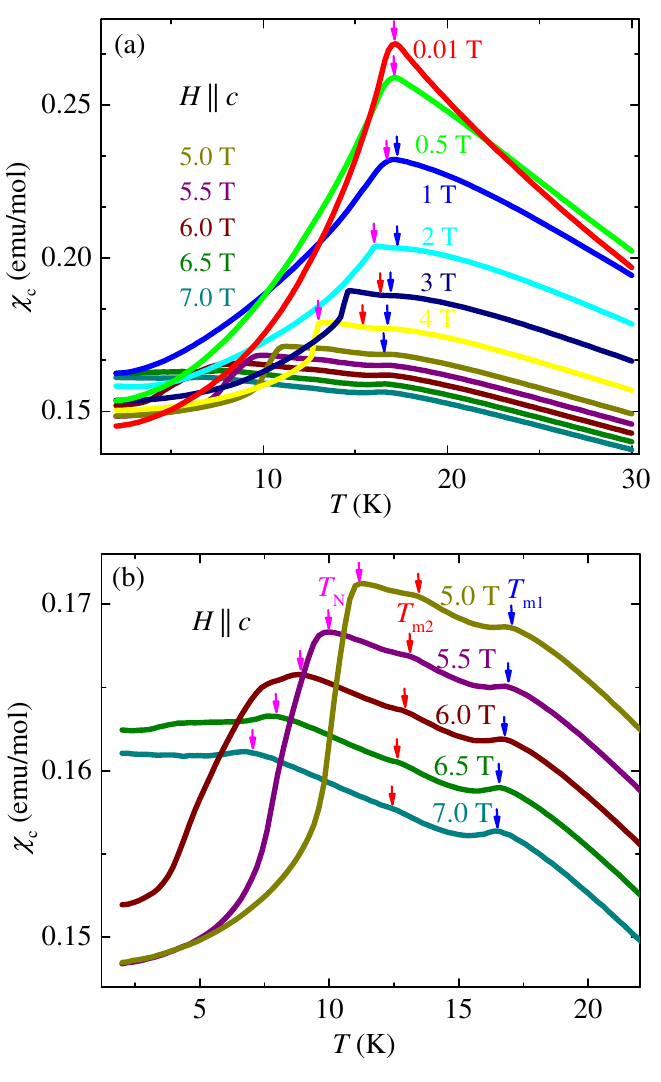}
	\caption{\label{PMT}(a) Temperature-dependent magnetic susceptibility of GdAuGe measured under different applied magnetic fields of $\mu_0$\textit{H} = 0.01, 0.5, 1, 2, 3, 4, 5, 5.5, 6.0, 6.5, and 7.0 T for \textit{H} $\parallel$ \textit{c}. (b) Magnified view of magnetic susceptibility curves at higher fields of $\mu_0$\textit{H} = 5.0, 5.5, 6.0, 6.5, and 7.0 T. The arrows are guide to various magnetic transitions such as AFM \textit{T}$_{N}$ (magenta) and field-induced anomalies \textit{T}$_{m1}$ (blue) and \textit{T}$_{m2}$ (red).}
\end{figure}

\begin{figure}
	\includegraphics[width=7.0cm, keepaspectratio]{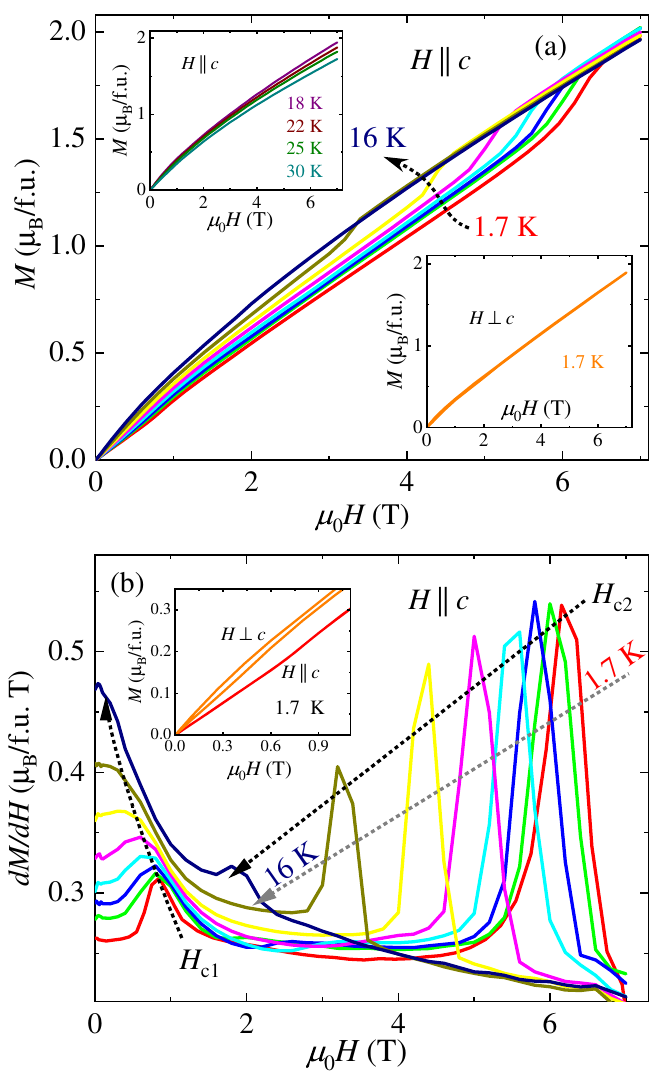}
	\caption{\label{MH}(a) Isothermal magnetization of GdAuGe at several different temperatures of \textit{T} = 1.7, 4, 6, 8, 10, 12, 14, and 16 K for \textit{H} $\parallel$ \textit{c}. Top inset of (a) shows the magnetic isotherms in temperature range of \textit{T} = 18$-$30 K.  Bottom inset of (a) presents the magnetic field dependence of magnetization measured at temperature \textit{T} = 1.7 K for \textit{H} $\perp$ \textit{c}. (b) Magnetic field dependence of differential magnetization at the various temperatures for \textit{H} $\parallel$ \textit{c}. The dotted arrows show the critical fields \textit{H}$_{c1}$ and \textit{H}$_{c2}$ at various temperatures corresponding to two MM transitions observed in GdAuGe. Inset of (b) shows a zoom view of low-field magnetizations at 1.7 K in both field-directions.}
\end{figure}

Next, we study the $\chi_c$(\textit{T}) behavior under different applied magnetic fields (0.01$-$7 T) along the \textit{c} axis. The data plotted under various magnetic fields are shown in Figs. \ref{PMT}(a) and \ref{PMT}(b). At very low field, $\mu_0$\textit{H}$_c$ $\sim$ 0.01 T, a very sharp peak, which is a typical characteristics of an AFM ordering can be observed at \textit{T}$_{N}$ $\sim$ 17.2 K (Fig. \ref{PMT}(a)). With further increase in field, this peak gets suppressed in magnitude and becomes broad along with shift towards low temperatures. Above 0.5 T, we observe the onset of field-induced anomalies in addition to an AFM transition. These anomalies shift towards low temperatures with an increase in the field. Fig. \ref{PMT}(b) presents the magnified view of these anomalies along with AFM transition at higher fields. The peaks of anomalies are marked by arrows to facilitate the view. It is interesting to note that AFM peak becomes almost flat at $\mu_0$\textit{H}$_c$ $\sim$ 7 T, and the curvature of $\chi$ vs. \textit{T} tends to approach the FM state through these field-induced anomalies in the system.

Further, we measured isothermal magnetization for \mbox{\textit{H} $\parallel$ \textit{c}} between 1.7 and 30 K with magnetic fields up to 7 T, as shown in Fig. \ref{MH}(a). They show monotonic increase in magnetization values with no sign of saturation. Magnetization value reaches $\sim$ 2.02 $\mu_B$/Gd at 7 T, which is much smaller than the value expected for free Gd$^{3+}$ ion. Magnetic isotherms further reveal the sudden change in slope at two critical fields below 18 K, indicating the emergence of two MM transitions in GdAuGe. It is noteworthy to mention here that we also measured in-plane magnetization (\textit{H} $\perp$ \textit{c}). But it did not reveal any MM transition besides the negligible hysteresis at low magnetic fields (see insets of Figs. \ref{MH}(a) and \ref{MH}(b)). Next, we calculate the MM critical fields at various temperatures using the maxima of field-dependent differential magnetization curves as shown in Fig. \ref{MH}(b). At 1.7 K, the critical fields are $\mu_0$\textit{H}$_{c1}$ $\sim$ 0.8 and $\mu_0$\textit{H}$_{c2}$ $\sim$ 6.2 T, which decrease with increasing temperature. This trend is in accordance with spin-flop transition expected in antiferromagnets \cite{RNiSi3} and has been observed in other Gd-based compounds such as Gd$_2$Te$_3$ \cite{Gd2Te3}. Above 16 K, MM transitions completely disappear and the compound enters into PM state.   

\begin{figure}
	\includegraphics[width=6.80cm, keepaspectratio]{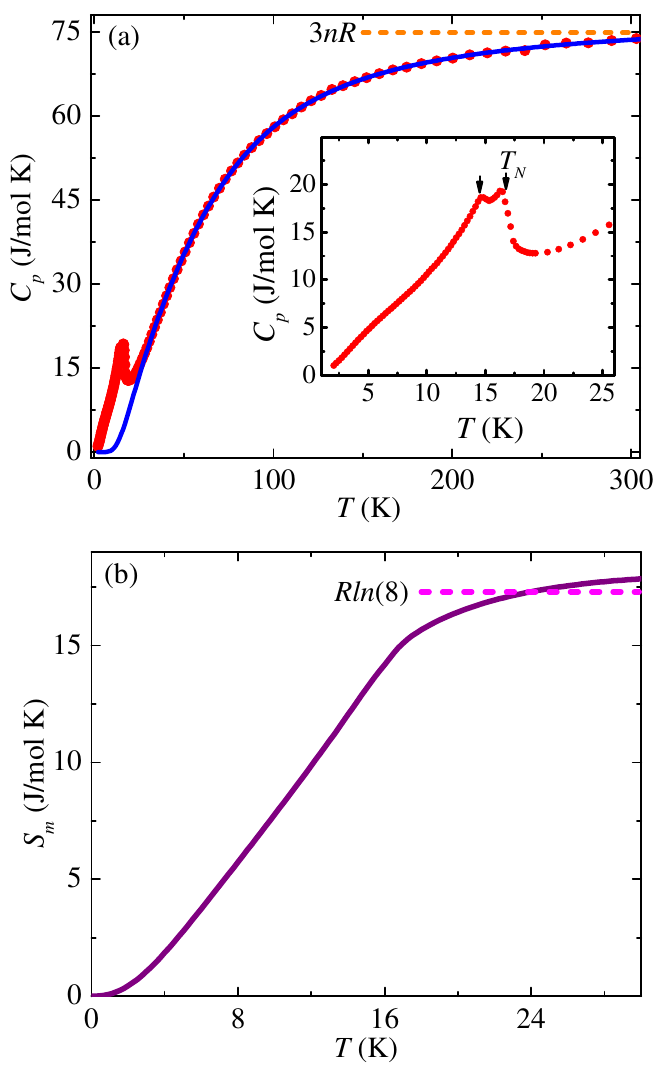}
	\caption{\label{HC}(a) Temperature-dependent heat capacity (\textit{C}$_p$) of GdAuGe single crystal. The solid blue line represents the Debye-Einstein Model fit to the experimental data. The inset shows a magnified view of the \textit{C}$_p$ behavior at low temperatures. (b) Magnetic entropy of GdAuGe as a function of temperature.}
\end{figure}

\begin{figure*}[!htp]
	\includegraphics[width=17.8cm, keepaspectratio]{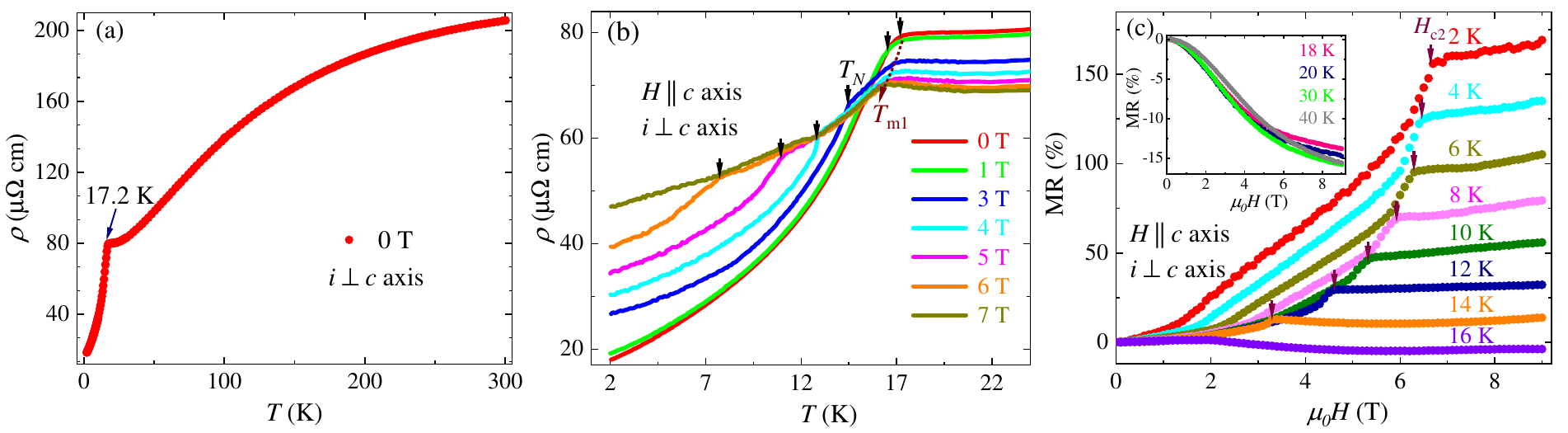}
	\caption{\label{Rho}Temperature-dependent electrical resistivity of GdAuGe single crystal in (a) without and (b) with magnetic field. The current is applied along \textit{ab} plane of crystal. Transverse magnetoresistance as a function of magnetic field at different temperatures (c) below \textit{T}$_N$ and inset above \textit{T}$_N$ for \textit{H} $\parallel$ \textit{c}. The arrows and dotted line mark different phase transitions.}
\end{figure*}

The origin of MM transitions in our data remains unclear like other rare-earth silver and gold germanides systems. Although, it is observed that a number of factors such as strong magnetocrystalline anisotropy, CEF effects, and competition between long-range FM and AFM interactions contribute to the MM transitions observed in rare earth compounds \cite{RAgGe_2004,RPtIn,Metamagnetism}. In the present case, magnetocrystalline anisotropy is small and CEF effects are minimal considering the fact that Gd$^{3+}$ ions are in symmetric $^8$\textit{S}$_{7/2}$ state \cite{RPtIn}. Recently, multiple MM transitions observed in CeRh$_3$Si$_2$ were explained using the Ising model, which generates the series of commensurate and incommensurate phases, leading to the metamagnetism like features \cite{CeRh3Si2}. The transition from a commensurate to incommensurate phase was reported in isostructural HoAuGe, which shows MM transition $\sim$ 0.4 T at 2 K \cite{HoAuGe}. Such possibility in GdAuGe is subject of future investigations.

\subsection{Heat capacity and entropy}
Figure \ref{HC}(a) shows the heat capacity (\textit{C}$_p$) of GdAuGe single crystal measured in \textit{T} range 2$-$300 K. A broad peak feature in low temperature \textit{C}$_p$ data near \textit{T}$_N$ is consistent with an AFM ordering observed from the magnetic measurements. The size of this peak ($\Delta$C$_p$) is  $\sim$ 7.2 J/mol K, which is almost half of the value predicted by mean-field theory for amplitude-modulated magnetic structure. A close zoom-in view of the peak shows that it is split into two parts at \textit{T}$_N$ = 16.8 and 14.9 K (see inset of Fig. \ref{HC}(a)). Two magnetic transitions in GdAuGe based on \textit{C}$_p$(\textit{T}) data were reported earlier in Ref. \cite{GdAuGe1996,RAuGe2023}. It is suggested to be associated with spin-reorientation processes in the compound \cite{GdAuGe1996}. Furthermore, we observe a broad hump around 6 K in low temperature \textit{C}$_p$ data. This kind of broad hump has been reported in other Gd based compounds, for example, in GdCu$_2$Si$_2$, it is observed at $\sim$ 3 K and is associated with emergence of (2\textit{J}+1)-fold degeneracy of multiplet in the ordered system \cite{Gd_Compounds,Gd_Compounds_1991}.

\begin{figure*}[t]
	\centering
	\includegraphics[width=17.4cm, keepaspectratio]{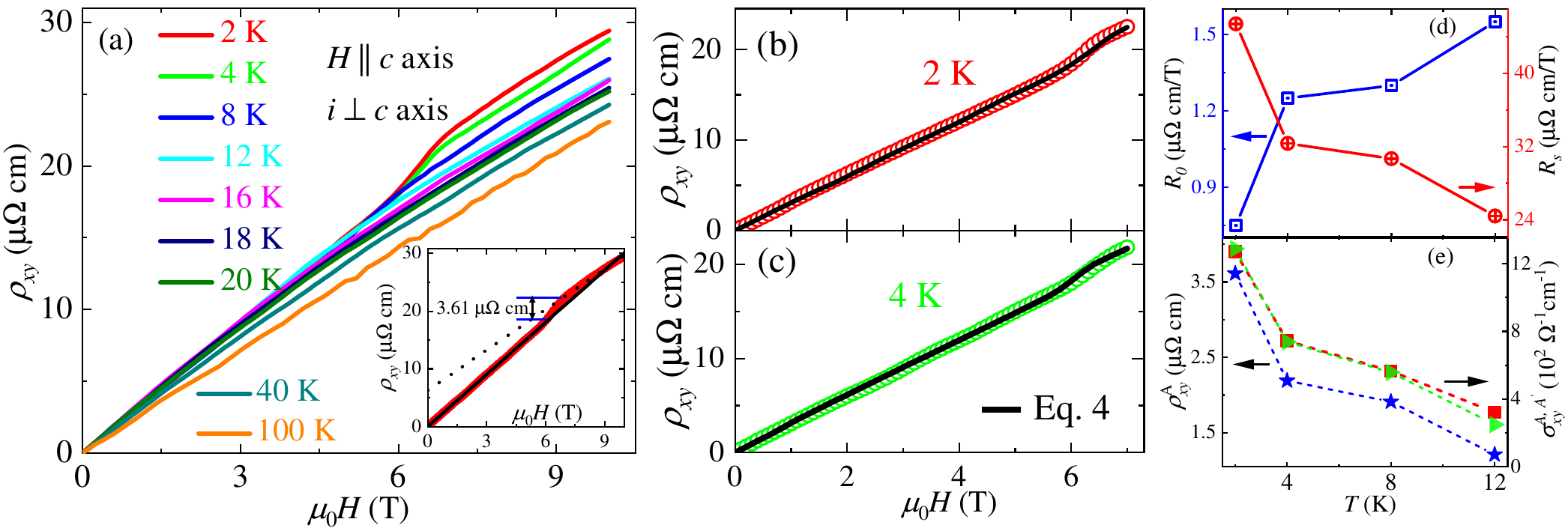}
	\caption{(a) The Hall resistivity measured at various temperatures from 2 to 100 K. The inset displays the magnitude of anomalous Hall resistivity at 2 K. In panels (b) and (c), the solid black line represents the fit of Eq. (\ref{AHE}) to the experimental data at 2 and 4 K. Panel (d) shows the temperature variation of ordinary and anomalous Hall coefficient. (e) The anomalous Hall resistivity $\rho_{xy}^A$ (\textcolor{blue}{$\displaystyle \star$}), and conductivity $\sigma_{xy}^{A}$ (\textcolor{red}{$\displaystyle \blacksquare $}) and $\sigma_{xy}^{A'}$ (\textcolor{green}{$\displaystyle \blacktriangleright$}) as a function of temperature.}
	\label{Hall}
\end{figure*}

At 300 K, \textit{C}$_p$ value approaches $\sim$ 73.74 J/mol K, which is within the Dulong-Petit limit. Attempts to determine the electronic specific-heat coefficient, $\gamma$ from low temperature \textit{C}$_p$ data fail due to the nonlinearity caused by the magnetic anomalies. \textit{C}$_p$ data above \textit{T}$_N$ can be well described by the following expression
\begin{equation}
C_p(T) = \gamma T+qC_{D}(T)+(1-q)C_{E}(T)
\label{Eq1}
\end{equation}
where $q$ is the weight factor, and $C_D$(\textit{T}) and $C_E$(\textit{T}) are Debye and Einstein \textit{C}$_p$ contributions, respectively, defined as
\begin{equation}
C_{D}(T)=9nR\left( \frac{T}{\Theta_D}\right)^3\int_{0}^{\Theta_D/T}\frac{x^4e^x}{(e^x-1)^2}dx
\end{equation}
and
\begin{equation}
C_{E}(T)=3nR\left( \frac{\Theta_E}{T}\right)^2\frac{e^{\Theta_E/T}}{(e^{\Theta_E/T}-1)^2}
\end{equation} 
where $\Theta_D$ and $\Theta_E$  are the Debye and Einstein temperatures, respectively. The values of various fitting parameters follow as, $\gamma$ = 3.9 mJ/mol K$^2$, $\Theta_D$ = 307, $\Theta_E$ = 90 K, and $q$ = 0.58. The magnetic entropy (shown in Fig. \ref{HC}(b)) is calculated using the formula, \textit{S}$_{m}= \int \frac{C_{m}}{T}dT$, where magnetic contribution, \textit{C}$_m$ is obtained by subtracting lattice part from the experimental data using Eq. (\ref{Eq1}). The entropy \textit{S}$_m$ released at \textit{T}$_N$ is slightly lower than the theoretical value \textit{S} = $R$\textit{ln}(2\textit{J}+1) = 17.3 J/mol K for Gd$^{3+}$ with \textit{J} = 7/2. The \textit{S}$_m$(\textit{T}) reaches \mbox{$R$\textit{ln}8} at 24 K and then saturates above 28 K. A slightly higher value of saturation entropy is due to the partial subtraction of phonon contribution \cite{EuPtAs}.

\subsection{Magnetotransport}

The electrical resistivity, $\rho$ as a function of \textit{T} measured along the \textit{ab} plane of crystal is shown in Fig \ref{Rho}(a). The investigated crystal shows the room temperature resistivity value of around 206 $\mu$$\Omega$ cm, and residual resistivity ratio ($\rho _{300 K}$/$\rho _{2 K}$) $\approx$ 11.33. It is comparable to the values reported in the literature for other Gd based ternary intermetallic compounds \cite{RPdSi,GdPdX}. The $\rho$(\textit{T}) exhibits a typical metal like behavior. It decreases systematically with decreasing \textit{T} until it registers a sharp drop in value near the magnetic transition temperature. The sharp drop at \textit{T}$_N$ = 17.2 K is the result of substantial reduction in the spin-disorder scattering and corroborates the results of magnetic and heat capacity measurements. Furthermore, we also measured the temperature-dependent electrical resistivity with magnetic field \textit{H} $\parallel$ \textit{c}, as shown in Fig. \ref{Rho}(b). As field strength increases, the AFM transition \textit{T}$_N$ shifts to lower temperatures, as marked with black arrows. Above 1 T, a second anomaly appears at \textit{T}$_{m1}$, which does not change so much with magnetic field. The values of \textit{T}$_{N}$ and \textit{T}$_{m1}$ are consistent with the $\chi_c$(\textit{T}) data as discussed above.

The transverse magnetoresistance (MR) measured for field configuration \textit{H} $\parallel$ \textit{c} in \textit{T} range 2$-$40 K, are shown in Fig. \ref{Rho}(c). The MR is positive for \textit{T} $\le$ 14 K and becomes negative at \textit{T} $\ge$ 16 K, as observed in AFM systems. In weak fields, it follows \textit{H}$^{1.3}$ field dependence and its value reaches only about 7\% at 0.8 T. For fields higher than 0.8 T, a weak anomaly is visible in MR data, thereafter, MR increases sublinearly. It reaches up to $\approx$ 123\% at 2 K until an onset of another anomaly at 6.2 T. In the vicinity of this anomaly, MR value suddenly jumps to $\approx$ 160\% and tends to saturate above 7 T. The anomalies observed in MR data in the vicinity of critical fields, where we observed the MM transitions in magnetic isotherms, clearly indicate that they are related to the MM transitions observed in the compound. The positive MR below \textit{T}$_N$ for an AFM phase is quite naturally expected, however large and positive MR due to the MM transitions in GdAuGe is in contrast to small and negative MR observed in several rare-earth compounds \cite{RAgGe_2004,RNi2Ge2}. Ideally, application of magnetic field reduces the electrical resistivity of ferromagnet and paramagnet, leading to negative MR. However, in the literature, numerous cases of this type of sudden enhancement in MR have been observed \cite{RAgGe_2004,Ce2Ni3Ge5,Ce2Ni3Ge5_2018,Tb5Si3,EuAg4As2}. The MR value of our crystal is almost twice the value of $\sim$ 82\% observed for TbAgGe crystals \cite{RAgGe_2004} and comparable to the value reported for EuAg$_4$As$_2$ single crystals \cite{EuAg4As2}. With increasing \textit{T}, sharp steps of increase in MR observed due to MM transitions gradually decreases and completely disappear above \textit{T}$_N$, leading to the negative values of MR. 

\begin{figure}[t]
	\centering
	\includegraphics[width=85mm,height=40mm]{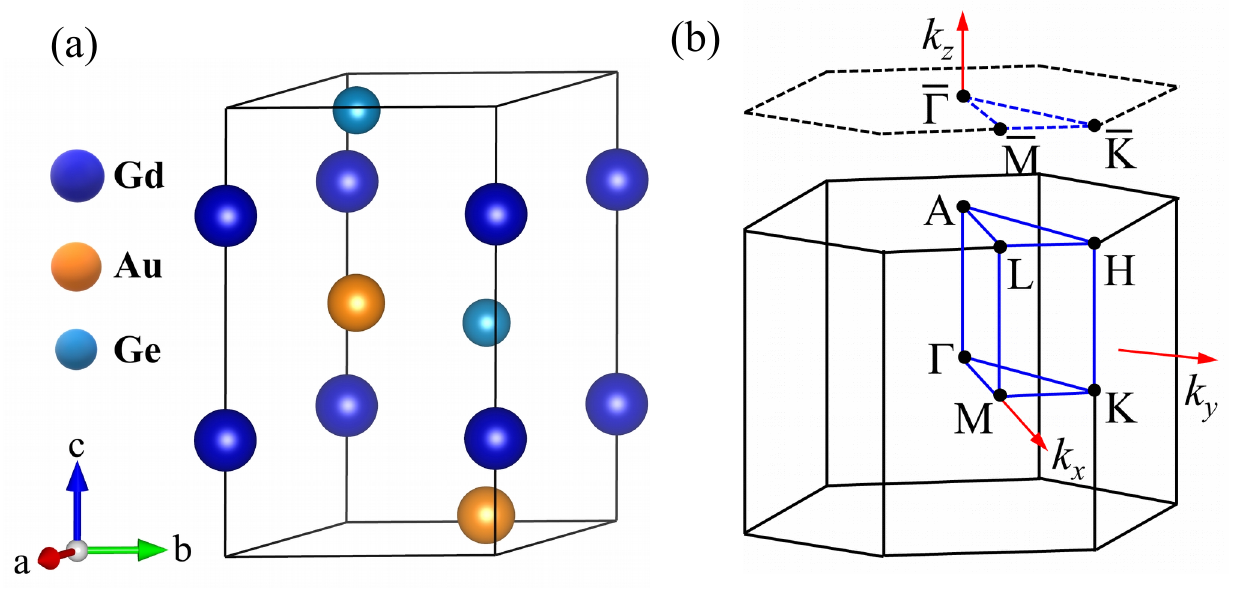}
	\caption{(a) The crystal structure of GdAuGe. (b) The irreducible Brillouin zone of the bulk along with the (001) projected surface.}
	\label{crystal}
\end{figure}

\begin{figure*}[t]
	\centering
	\includegraphics[width=15.40cm, keepaspectratio]{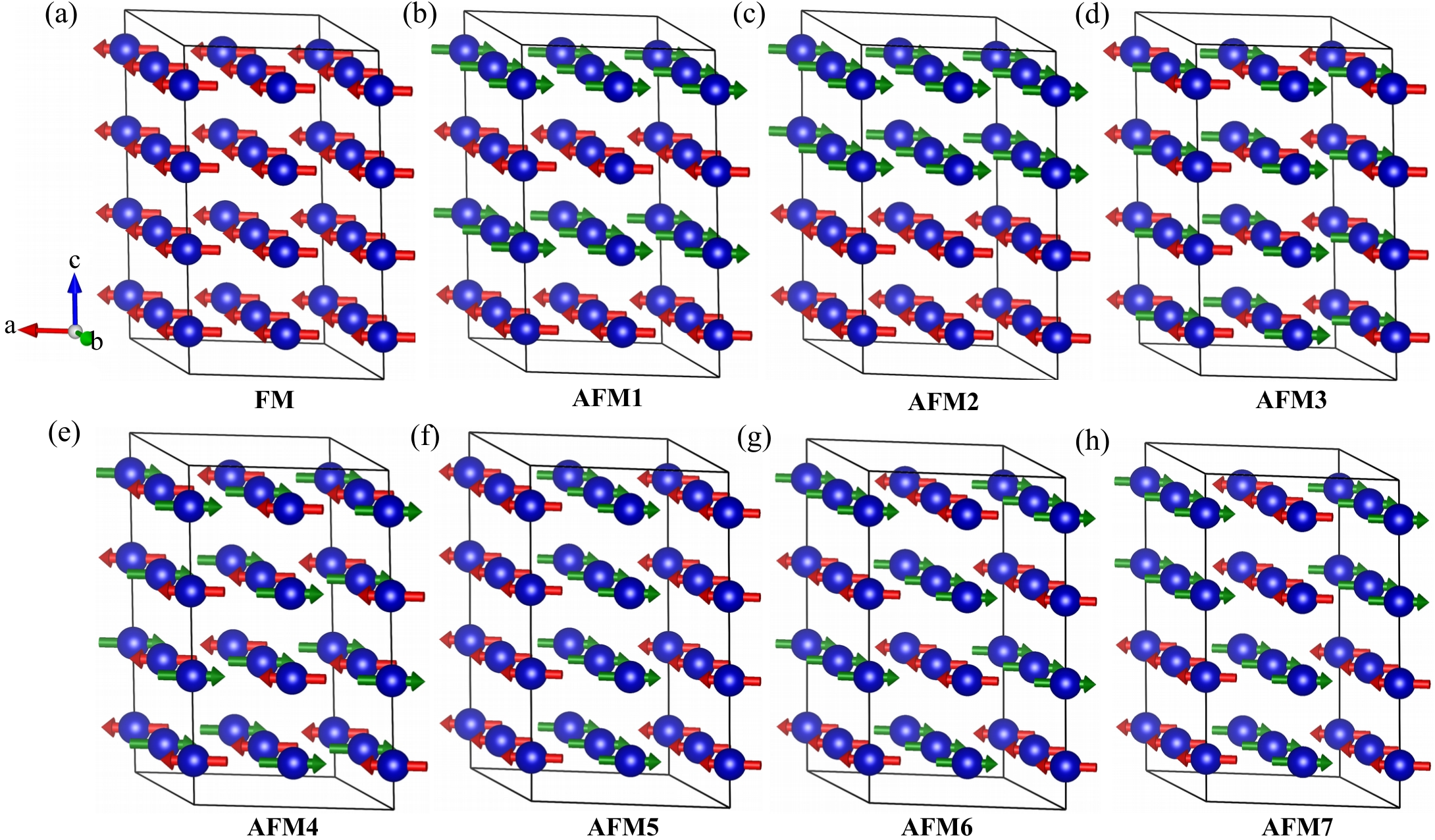}
	\caption{(a)$-$(h) FM and AFM configurations for 2 $\times$ 2 $\times$ 2 supercell with Gd spins. Here, AFM1, AFM3, AFM4 are \textit{A}-, \textit{C}-, \textit{G}-type, respectively, whereas other configurations are stripe-type AFM. Red and green arrows denote the spin-up and spin-down, respectively.}
	\label{afm}
\end{figure*}

Figure \ref{Hall}(a) displays the Hall resistivity ($\rho_{xy}$) of single-crystalline GdAuGe, measured within the \textit{ab} plane over a \textit{T} range of 2 to 100 K. The $\rho_{xy}$ increases continuously with an increasing magnetic field in a slightly nonlinear manner and its value remains positive throughout the temperature range, indicating that holes are the majority charge carriers. Moreover, we estimate the carrier concentration and mobility to be approximately 2.69 $\times$ 10$^{20}$ cm$^{-3}$ and 167 cm$^{2}$ V$^{-1}$ s$^{-1}$, respectively, by obtaining the slope from the linear fit of the 100 K dataset. At low temperatures (below 16 K), the $\rho_{xy}$ exhibits a step-like increase around the critical magnetic field range, where we observed the signature of MM transition in magnetization and MR data. Considering that the step-like feature in $\rho_{xy}$ is a part of the anomalous Hall resistivity ($\rho_{xy}^A$). To calculate the magnitude of $\rho_{xy}^A$, we adopt the method used in Ref. \cite{RMn6Ge6}, as shown in the inset of Fig. \ref{Hall}(a). The $\rho_{xy}^A$ at 2 K is around 3.61 $\mu\Omega$ cm, and its magnitude decreases with increasing \textit{T}, reaching $\sim$ 1.2 $\mu\Omega$ cm at 12 K (see the Fig. \ref{Hall}(e)). In general, the total Hall resistivity including the $\rho_{xy}^A$ term is given by the following expression
\begin{equation}
\rho_{xy} = \rho_{xy}^O + \rho_{xy}^A= R_0H+R_s\mu_0M,
\label{AHE}
\end{equation}
 where \textit{R}$_{0}$ and \textit{R}$_{s}$ are the ordinary and anomalous Hall coefficients, respectively and \textit{M}(\textit{H}) is isothermal magnetization data as a function of field. In our data, it is difficult to separate $\rho_{xy}^O$ and $\rho_{xy}^A$ contributions as the magnetic moments do not saturate up to field strength 7 T. Therefore, we have simulated the experimental data up to $\mu_0H = $ 7 T using Eq. (\ref{AHE}). The results for 2 and 4 K data set are displayed in Figs. \ref{Hall}(b) and \ref{Hall}(c), respectively. The estimated values of $R_0$ and $R_s$ from simulated data are presented in Fig. \ref{Hall}(d). The values of $R_s$ are significantly larger than $R_0$. Next, we present the anomalous Hall conductivity (AHC),  $\sigma_{xy}^A$ = $\rho_{xy}^A$/($\rho_{xx}^2 + \rho_{xy}^2$), in Fig. \ref{Hall}(e), and the AHC decreases with the increasing temperatures. At 2 K, its value is about 1270 $\Omega$$^{-1}$ cm$^{-1}$, which is of the same order as reported for AFM topological systems DyPtBi \cite{DyPtBi} and TbPtBi \cite{TbPtBi}. To further check the consistency of calculated AHC, we have estimated the AHC ($\sigma_{xy}^{A'}$) using the $R_s$ and change in magnetization value around the MM transition. The obtained values of $\sigma_{xy}^{A'}$ are quite close to that directly calculated from $\rho_{xy}$(\textit{H}) curves (see Fig. \ref{Hall}(e)).

\subsection{Electronic structure}

\begin{figure}[t]
	\centering
	\includegraphics[width=88mm,height=76mm]{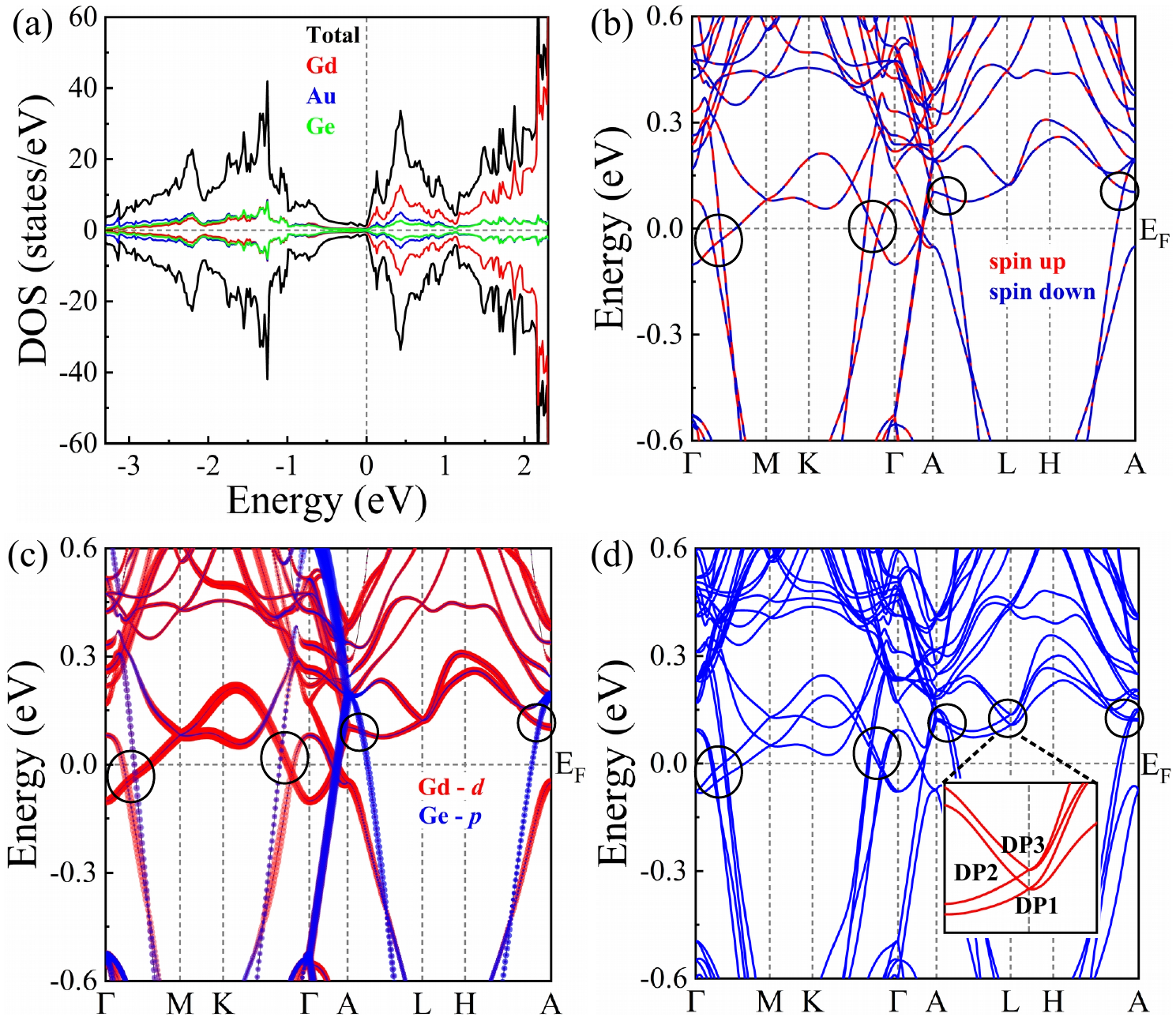}
	\caption{(a) Total and projected density of states of GdAuGe. (b) Electronic band structure along $\Gamma$-$M$-$K$-$\Gamma$-$A$-$L$-$H$-$A$ path without SOC. (c) The orbital-decomposed electronic band structure without SOC. (d) The electronic band structure with SOC. Inset shows the Dirac points DP1, DP2 and DP3.}
	\label{bands}
\end{figure}

\begin{figure}[t]
	\centering
	\includegraphics[width=85mm,keepaspectratio]{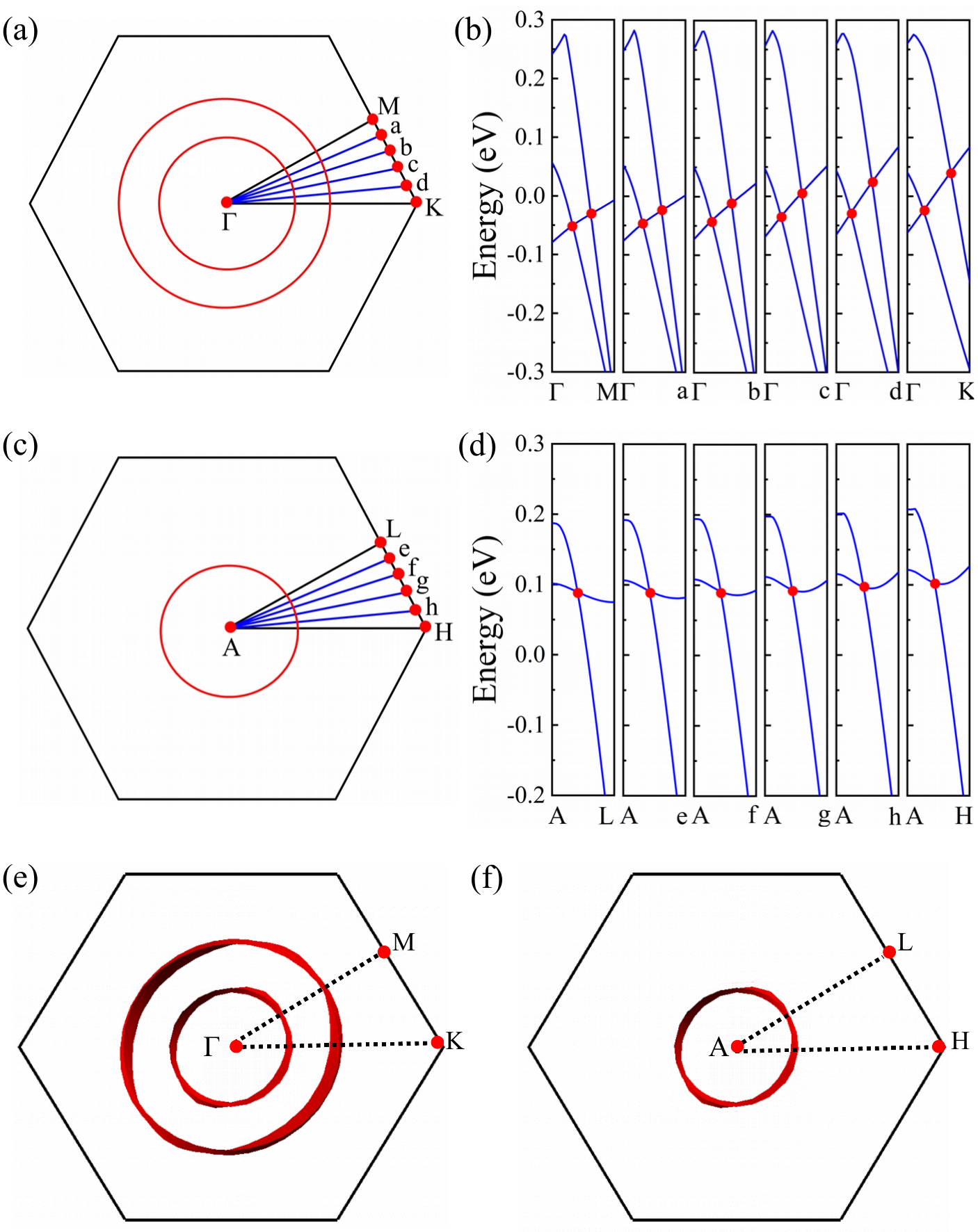}
	\caption{(a) and (c) The illustration of the nodal line, where $a$, $b$, $c$, $d$ are equally spaced points between $M$ and $K$ along \textit{k}$_z$ = 0 plane, and $e$, $f$, $g$, $h$ between $L$ and $H$ along \textit{k}$_z$ = 0.5 plane, respectively. (b) and (d) Electronic band structures along the \textit{k}-paths as indicated in (a) and (c), respectively. Iso-energy Fermi contours along (e) \textit{k}$_z$ = 0 and (f) \textit{k}$_z$ = 0.5 planes, which show the nodal lines.}
	\label{nodal}
\end{figure}

The unit cell of GdAuGe consists of six atoms, with Gd, Au, and Ge atoms occupying Wyckoff positions 2\textit{a}, 2\textit{b}, and 2\textit{b}, respectively. As shown in Fig. \ref{crystal}(a), it crystallizes in a hexagonal structure with the space group $P$6$_3$${mc}$ (186). Along with three vertical mirror planes $\widetilde{M}$$_{x\bar{y}}$ = \{\textit{M}$_{x\bar{y}}$$|$00$\frac{1}{2}$\}, $\widetilde{M}$$_{2xy}$ = \{\textit{M}$_{2xy}$$|$00$\frac{1}{2}$\}, and \textit{M}$_y$, the structure has threefold rotational symmetry, \textit{C}$_{3z}$ and twofold screw rotational symmetry, \textit{S}$_{2z}$ = \{\textit{C}$_{2z}$$|$00$\frac{1}{2}$\}. In Fig. \ref{crystal}(b), we display the (001) surface  Brillouin zone (BZ) beside the bulk BZ. In order to investigate the possible magnetic configurations, we have examined the FM and seven AFM (including \textit{A}-, \textit{C}-, \textit{G}- and stripe-type) spin configurations with 2 $\times$ 2 $\times$ 2 supercell. The possible magnetic configurations are shown in Fig. \ref{afm}. The calculated ground state energy for each configuration is presented in Table \ref{table}. From the Table \ref{table}, it can be seen that the AFM5 configuration yields the lowest energy. Here, AFM5 configuration exhibits the AFM coupling along the \textit{a} axis, whereas FM coupling along the \textit{b} and \textit{c} axes. To further confirm the spin orientations in the AFM5 case, we have calculated the ground state energies along different spin alignments such as [001], [010], [100], [011], [101], [110] and [111]. The computed ground state energy differences are given in Table \ref{SOC_table}. The minimum ground state energy is observed for the [100] spin configuration. A similar AFM5 magnetic structure has also been reported in the isostructural compounds RAuGe (R = Tb$-$Er), where magnetic moments are inclined with respect to the \textit{c} axis \cite{HoAuGe,Er_HoAuGe,RAuGe}. This inclination angle of magnetic moment decreases with increase in number of 4\textit{f} electrons. For example, TbAuGe magnetic moment is inclined at an angle of 65$^\circ$ to the \textit{c} axis, while ErAuGe magnetic moment is along the \textit{c} axis. Following the trend of the magnetic structure of isostructural RAuGe compounds, the magnetic moments of GdAuGe are likely to be aligned along the \textit{ab} plane, as suggested by our DFT calculations. However, it cannot be completely ascertained, as our experimental data indicate that moments are preferably aligned along the \textit{c} axis. Further, the CEF effects are absent in GdAuGe, unlike isostructural RAuGe compounds, which could affect the orientation of magnetic moments \cite{Er_HoAuGe}. Our DFT results are valid for \textit{T} = 0 K. Furthermore, we have used a fixed value of U (= 6 eV) in the absence of experimentally determined U value. Thus, correlation effects are not taken care of appropriately. These inherent limitations might be responsible for the difference between the theoretically predicted magnetic structure and the magnetic measurements. To determine the precise orientation of Gd spins within the GdAuGe and to resolve the discrepancy between our theoretical calculations and experimental observations, further investigations are required, especially using microscopic techniques.

\begin{table}[t]
	\caption{Calculated energies of different magnetic configurations (in meV) with the reference energy considered to be 0 meV.}
	\begin{tabular}{c c c c}
		\hline\hline
		Configuration & Energy (meV) & Configuration & Energy (meV) \\ [0.5ex]  
		\hline
		 FM &11.93& AFM4 & 3.59 \\[0.5ex]  
		AFM1&5.72 & AFM5 & 0.00 \\ [0.5ex]  
		AFM2&5.71 & AFM6 & 3.59 \\ [0.5ex]  
	    AFM3&0.02 & AFM7 & 2.79 \\ [0.5ex]  

		\hline\hline
	\end{tabular}
	\label{table}
\end{table}

\begin{table}[t]
\caption{Calculated energies of different spin configurations in AFM5 case with the reference energy considered to be 0 $\mu$eV.}
\begin{tabular}{ c c c c c c c c}
\hline\hline
Configuration & [001] &~ [010] & ~[100] & ~[011] &~ [101] &~ [110] & ~[111] \\ [0.5ex] 
\hline 
Energy ($\mu$eV) & 49.79 & 21.56 & 0.00 & 34.79 & 24.12 & 20.11 & 15.09 \\ [1ex] 
\hline\hline
\end{tabular}
\label{SOC_table}
\end{table}

\begin{figure}
	\includegraphics[width=7.3cm, keepaspectratio]{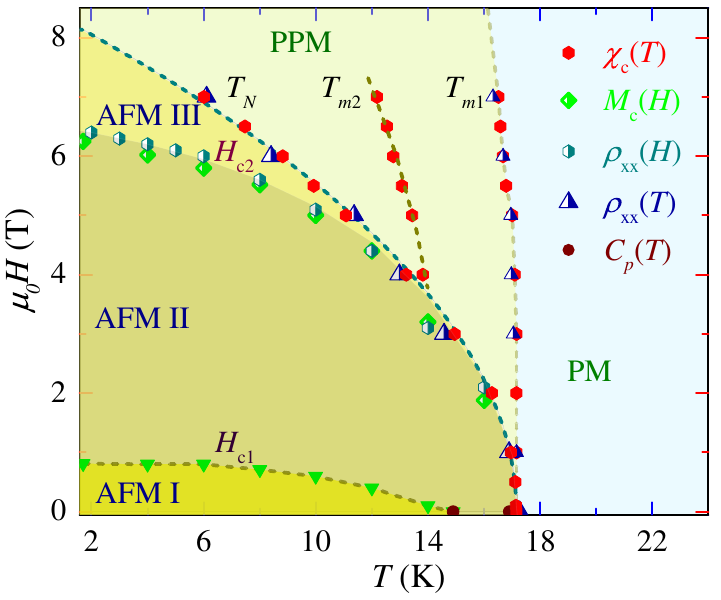}   	
	\caption{\label{Phase}The \textit{H}-\textit{T} phase diagram of the GdAuGe when magnetic field is applied along the \textit{c} axis. Denotations are mentioned in main text. The dark cyan dot line illustrates the fit of molecular field theory equation \textit{H}=\textit{H}$_0$[1$-$\textit{T}$_N$(\textit{H})/\textit{T}$_N$(\textit{H} = 0)]$^{1/2}$ to the experimental data. Dotted gray lines are just guides to the eyes.}
\end{figure}

Furthermore, the total density of states (DOS) and projected density of states (PDOS) were calculated for AFM5 case, to illustrate the behavior of Gd, Au and Ge elements, and the results are displayed in Fig. \ref{bands}(a). The valence band region is equally contributed by Gd, Au and Ge atoms, whereas the conduction region is dominated by Gd in both spin channels. Moreover, GdAuGe has a small DOS at the Fermi level, which confirms the semimetallic nature of the compound. We have also investigated the electronic band structure properties. Figure \ref{bands}(b) shows the electronic band structure with the spin-up (in red color) and the spin-down (in blue color) channels. The electronic band structure exhibits some band crossing points near the Fermi level along the \textit{k}$_z$ = 0 as well as \textit{k}$_z$ = 0.5 plane, which might lead to the nodal line. To determine the non-trivial nature of these bands, we calculated the orbital decomposed band structure (Fig. \ref{bands}(c)) and it infers that Gd-$d$ and Ge-$p$ states are main contributors to the band crossing points. From the Fig. \ref{bands}(c), we observe two crossing points along \textit{k}$_z$ = 0 plane. The band inversion between Gd-$d$ and Ge-$p$ in one crossing point can be seen, which reflects the non-trivial nature of crossing points, whereas another crossing point lacks the band inversion and shows the trivial nature of crossings. Similarly, the band inversion can also be seen along \textit{k}$_z$ = 0.5 plane. Notably, each band along the \textit{k}$_z$ = 0.5 plane is twofold degenerate due to the anticommutation relation between \textit{M}$_{y}$ and \textit{S}$_{2z}$ symmetries \cite{twofold}, which show the four-fold degeneracy in bands at the crossing point and hints towards the presence of a Dirac nodal line. To analyze these band crossings, we have performed a detailed calculation of the band structures along $\Gamma$-$M/a/b/c/d/K$ paths (see Fig. \ref{nodal}(a)) as well as $A$-$L/e/f/g/h/H$ paths (see Fig. \ref{nodal}(c)) and found that the Dirac-type band crossings appeared in all the above-mentioned paths (see Figs. \ref{nodal}(b) and \ref{nodal}(d)) reflecting the occurrence of two $\Gamma$-centered nodal rings protected by \textit{M}$_{y}$ symmetry, and one $A$-centered Dirac nodal ring protected by \textit{M}$_{y}$ and \textit{S}$_{2z}$ symmetries. Furthermore, we have confirmed the presence of nodal lines through iso-energy Fermi contours and shown in Figs. \ref{nodal}(e) and \ref{nodal}(f). With the inclusion of SOC, we can see the band opening at the crossing points, which is shown in Fig. \ref{bands}(d). In addition, there exist multiple Dirac points (DP1, DP2, and DP3) along $A$-$L$ path, which is shown in the inset of Fig. \ref{bands}(d). The Dirac points DP1, DP2, and DP3 are generated by \textit{M}$_{y}$ and non-symmorphic (\textit{S}$_{2z}$) symmetries.

\subsection{\textit{H}-\textit{T} phase diagram}

Based on the experimental data presented above, we have constructed the \textit{H}$-$\textit{T} phase diagram for \textit{H} $\parallel$ \textit{c}, which is depicted in Fig. \ref{Phase}. The phase line boundaries are calculated using the peak positions of derivatives of the  $\chi_c$(\textit{T}), \textit{M}(\textit{H}), $\rho_{xx}$(\textit{H}), and $\rho_{xx}$(\textit{T}) data. The resulting phase diagram shows four distinct regions in the magnetically ordered state. The first region, labeled AFM I, corresponds to AFM phase. The magnetic structure in this region below \textit{T}$_N$ is collinear AFM at low fields as evidenced by our electronic band structure calculations (see Fig. \ref{afm}(f)) and experimentally observed value of $\chi_c$(1.7 K)/$\chi_c$(\textit{T}$_N$) $\approx$ 0.5 (refer to Fig. \ref{MT}(a)). As the strength of field increases, we move into region AFM II. In this region, AFM spins tend to align along the direction of external magnetic field, and get partially flopped above the critical field \textit{H}$_{c1}$, which is deduced from the \textit{M}(\textit{H}) measurements. At \textit{H}$_{c1}$, the first spin-flop transition occurs, followed by the second spin-flop transition at the critical field \textit{H}$_{c2}$. After \textit{H}$_{c2}$ and below the mean field fitting line, the system is in region AFM III, which likely corresponds to an incommensurate magnetic structure. Both \textit{H}$_{c1}$ and \textit{H}$_{c2}$ decrease with increasing \textit{T}. The phase boundary of \textit{H}$_{c2}$ nearly overlap with that of \textit{T}$_N$ in the \textit{T} range 12$-$16 K. However, below 12 K, it is well separated from \textit{T}$_N$ and its values, derived from the magnetization and transport measurements, are in a good agreement. Such complex magnetic phases are also observed in the isostructural RAuGe (Tb$-$Er) \cite{Er_HoAuGe,HoAuGe,RAuGe}. We further note that \textit{T}$_N$ shifts towards low temperatures with increasing field like \textit{H}$_{c1}$ and \textit{H}$_{c2}$. Its trend is in well accord with that predicted by the molecular field theory equation \textit{H} = \textit{H}$_0$[1$-$\textit{T}$_N$(\textit{H})/\textit{T}$_N$(\textit{H} = 0)]$^{1/2}$, where \textit{H}$_0$ is the critical field strength required to completely destroy the AFM phase transition \cite{Magnetism}. The fit of this equation yields \textit{H}$_0$ = 8.57 T and \textit{T}$_N$(\textit{H} = 0) = 17.15 K. Further increase in field drives the system from region AFM III to PPM, which we refer to as the polarized paramagnetic (PPM) phase, where a large part of the magnetic moments is aligned along the field. It displays two field-induced anomalies, \textit{T}$_{m1}$ and \textit{T}$_{m2}$, in the vicinity of \textit{T}$_N$. These anomalies show small variation in their temperatures with increasing fields. As the temperature surpasses \textit{T}$_{m1}$, the compound transitions to the PM state.

\section{Conclusions}

We have investigated the magnetic, thermodynamic and magnetotransport properties of GdAuGe single crystals grown using Bi flux. The magnetic susceptibility measurements for field configuration \textit{H} $\parallel$ \textit{c} and \textit{H} $\perp$ \textit{c} revealed the AFM ground state in the compound with \textit{T}$_N$ = 17.2 K. The anomalies observed near \textit{T}$_N$ in the heat capacity data and a sharp drop in electrical resistivity data below \textit{T}$_N$ further confirmed the AFM ordering in the compound. The magnetization data for \textit{H} $\parallel$ \textit{c} showed two successive MM transitions at \textit{T} = 1.7 K with critical fields of $\sim$ 0.82 and 6.2 T. The magnetotransport data recorded for \textit{H} $\parallel$ \textit{c} near the critical fields of MM transitions was observed to show unexpectedly positive and large values of transverse MR (169\% at 9 T and 2 K) for temperatures less than \textit{T}$_N$. At higher temperatures, the MR decreases and becomes negative in the PM regime. A large anomalous Hall conductivity $\sim$ 1270 $\Omega$$^{-1}$ cm$^{-1}$ was observed near \textit{H}$_{c2}$ at 2 K. The phase-diagram in \textit{H} $\parallel$ \textit{c} vs. \textit{T} plane was constructed from the magnetization and magnetotransport measurements, which unveiled multiple magnetic phase transitions including a collinear AFM ground state, two successive spin-flop transitions and a polarized paramagnetic state corresponding to the two field-induced magnetic anomalies. The electronic band structure analysis shows the presence of two nodal rings along \textit{k}$_z$ = 0 plane and a Dirac nodal ring along \textit{k}$_z$ = 0.5 plane, which makes GdAuGe a Dirac nodal-line semimetal. 

\section{Acknowledgment}

We acknowledge IIT Kanpur and the Department of Science and Technology, India, [Order No. DST/NM/TUE/QM-06/2019 (G)] for financial support. J.S and V.K. acknowledge the National Supercomputing Mission (NSM) for providing computing resources of ‘PARAM SEVA’ at IIT, Hyderabad. V.K. would like to acknowledge DST-FIST (SR/FST/PSI-215/2016) for the financial support. J.S. was supported through a CSIR scholarship. K.S. and D.K. acknowledge financial support from the National Science Centre (Poland) under Research Grant No. 2021/41/B/ST3/01141.  

\bibliography{Reference_GdAuGe}

\end{document}